\documentclass[letterpaper, 10 pt, conference]{ieeeconf}  

\IEEEoverridecommandlockouts                              

\usepackage{graphics} 
\usepackage[ruled]{algorithm2e}
\usepackage{graphicx}
\usepackage{multirow}
\usepackage{url}
\usepackage{booktabs}
\usepackage{float}  
\usepackage{subfigure}  
\usepackage{color}
\usepackage{siunitx}
\usepackage{amssymb}
\usepackage{amsmath}
\usepackage{listings}
\usepackage{xcolor}
\newtheorem{theorem}{Definition}

\newcommand{\tool}{\textsc{RoadGen}\xspace}
\newcommand{\todo}[1]{\textcolor{black}{#1}}

\title{\LARGE \bf
\tool: Generating Road Scenarios for Autonomous Vehicle Testing
}

\author{Fan Yang$^{*\dagger}$, You Lu$^{*\dagger}$, Bihuan Chen$^{*\ddagger}$, Peng Qin$^{*}$, and Xin Peng$^{*}$
\thanks{This work was supported in part by the National Key R\&D Program of China (2021ZD0112903).}
\thanks{$^{\dagger}$Equal Contribution.}
\thanks{$^{\ddagger}$Corresponding Author.}
\thanks{$^{*}$With the School of Computer Science, Fudan University, Shanghai, China {\tt\small \{yangfan\_, bhchen, pengxin\}@fudan.edu.cn, \{luy22, pqin22\}@m.fudan.edu.cn}}%
}

\begin{document}

\maketitle
\thispagestyle{empty}
\pagestyle{empty}

\begin{abstract}
With the rapid development of autonomous vehicles, there is an increasing demand for scenario-based testing~to simulate diverse driving scenarios. However,~as~the~base of any driving scenarios, road scenarios (e.g., road topology~and~geometry) have received little attention by the literature. Despite~several advances, they either generate basic~road~components~without a complete road network,~or~generate~a~complete~road~network but with simple road components.~The~resulting road~scenarios lack diversity in both topology~and~geometry. To address this problem, we propose \tool to systematically generate diverse road scenarios. The key idea~is~to~connect eight~types~of parameterized road components to form road scenarios~with high diversity in topology and geometry. Our evaluation has demonstrated the effectiveness and usefulness of \tool in generating diverse road scenarios for simulation.
\end{abstract}


\section{Introduction}

Autonomous vehicles have been widely developed~in~the last decades due to their impact on automotive transportation and their benefit to society (e.g., reducing vehicle~collisions, and providing personal mobility to disabled people)~\cite{paden2016survey}. It is important to ensure the safety and reliability of autonomous vehicles for world-wide adoption. Therefore, on-road testing is widely used by leading companies.~However,~autonomous vehicles would have to be driven more than~11~billion miles to demonstrate with 95\% confidence that they are 20\% safer than human drivers~\cite{kalra2016driving}. It is expensive for on-road testing~to achieve this goal, and it is also impossible for on-road testing to test corner cases or dangerous situations.

To this end, scenario-based testing~\cite{riedmaier2020survey, zhong2021survey}~is~also~widely used by leading companies to simulate diverse driving scenarios.~As of February 2021, Waymo's autonomous vehicles have been tested with over 15 billion miles of simulated~driving~\cite{waymo}. Various approaches have been developed to generate driving scenarios~\cite{Zhang2023, Ding2023, schutt2023}. However, they mainly~focus on the diversity in vehicle/pedestrian behaviors and weather conditions, but overlook the diversity in roads~\cite{zhong2021survey, schutt2023}.~Several recent advances have been proposed~on~generating~road scenarios~\cite{zhou2023, rietsch2022driver, tang2022automatic, paranjape2020modular}. However, they either generate basic road components (e.g., highway interchanges) without a complete road network \cite{zhou2023, rietsch2022driver}, or build a complete~road network but with simple road components (i.e., straight~roads and junctions) \cite{tang2022automatic, paranjape2020modular}. Therefore, their generated road~scenarios lack diversity in both topology and geometry.

To address the problem, we propose \tool to systematically generate diverse road scenarios. First, we define~and implement eight types of basic road components. Each road component is parameterized to reflect its geometry diversity. Second, we connect road components to form road scenarios. This process is guided by favoring the selection~of~least~used road components so as to ensure the geometry diversity of road scenarios. Third, we remove duplicated road scenarios that have the same topology to ensure the topology diversity of road scenarios. Finally, we convert the generated road~scenarios into high-precision (HD) map files and 3D scene files, which can be used for joint simulation by simulators.

To evaluate the effectiveness of \tool, we use \tool to generate road scenarios that have 4, 5, 6, 7 and~8~road components. Our results have demonstrated that \tool can generate more diverse road scenarios than a baseline~approach that randomly selects and connects road components. Furthermore, to evaluate the usefulness~of~\tool, we sample road scenarios and convert them into HD map files and 3D scene files for joint simulation on SORA-SVL \cite{Huai2023} with Apollo 8.0 \cite{apollo}. Our results have~indicated that over 92\% of the road scenarios can be useful for joint simulation.

In summary, this work makes the following contributions.
\begin{itemize}
\item We define and implement eight types of parameterized basic road components.
\item We propose a guided approach to connect road components to generate diverse road scenarios.
\item We conduct experiments to demonstrate the effectiveness and usefulness of \tool, and build a dataset of road scenarios for simulation testing.
\end{itemize}

\section{Related Work and Problem Statement}

\subsection{Related Work}

Scenarios~\cite{Ulbrich2015, Menzel2018} are important for developing~and~testing autonomous vehicles. To represent scenarios, Bagschik~et al. \cite{Bagschik2018} develop a 5-layer model, including road-level (layer~1), traffic infrastructure (layer 2), manipulation~of~layer~1~and 2 (layer 3), objects (layer 4), and environment (layer~5), while Scholtes et al.~\cite{Scholtes2021} extend it by adding digital information as layer 6. To generate scenarios, various approaches have been developed~\cite{Sun2022, cai2022survey, Zhang2023, Ding2023, schutt2023}, e.g., extracting scenarios from driving data~\cite{kerber2020clustering, balasubramanian2021traffic} or crash data~\cite{gambi2019generating, bashetty2020deepcrashtest, esenturk2021analyzing, zhang2023building},~and searching scenarios by evolutionary algorithms~\cite{wang2021advsim, klischat2019generating, tian2022mosat} or combinatorial interaction testing~\cite{rocklage2017automated}.

However, recent surveys~\cite{zhong2021survey, schutt2023} show that most scenario generation approaches have focused on scenarios at layer 4 and 5 by manipulating vehicles, pedestrians and~weather~conditions, while little attention has been paid on road topology and geometry, and traffic signs (i.e., layer 1, 2 and 3),~which serve as the base of any scenario. Zhou et al.~\cite{zhou2023} propose~a model-driven method to generate highway interchanges that are one basic component in road networks.~Rietsch~et~al.~\cite{rietsch2022driver} also attempt to generate basic components, i.e., roundabout, intersections, highway entry, drive and exit. \todo{Differently,~our work aims to compose road networks based on basic components.} Tang et al.~\cite{tang2022automatic} first extract junction features from~HD maps, and then build road networks~by connecting~the~junctions in a grid layout. Paranjape et al.~\cite{paranjape2020modular} generate road~networks with different road sizes and intersections. \todo{However, these approaches only consider straight roads and junctions, and thus the generated road networks lack diversity.}

\subsection{Problem Statement}

This work is focused on the road-level (layer 1) scenario~of the 5-layer model~\cite{Bagschik2018}, which describes the topology~and~geometry of road scenarios. Specifically, the topology~of~a~road scenario can be characterized by how different road~components (e.g., straight road, curve road, fork road,~and~intersection) are connected together. The geometry of a road~scenario can be characterized by factors like~the~number~of~lanes~and the type of lane markings. The diversity of road scenarios in both topology~and~geometry is important for developing and testing autonomous vehicles. Therefore, our problem can be stated as \textit{how to systematically generate road~scenarios~such that they have diverse topology~and~geometry.}


\section{Methodology}

\begin{figure}[!t]
    \includegraphics[width=0.99\linewidth]{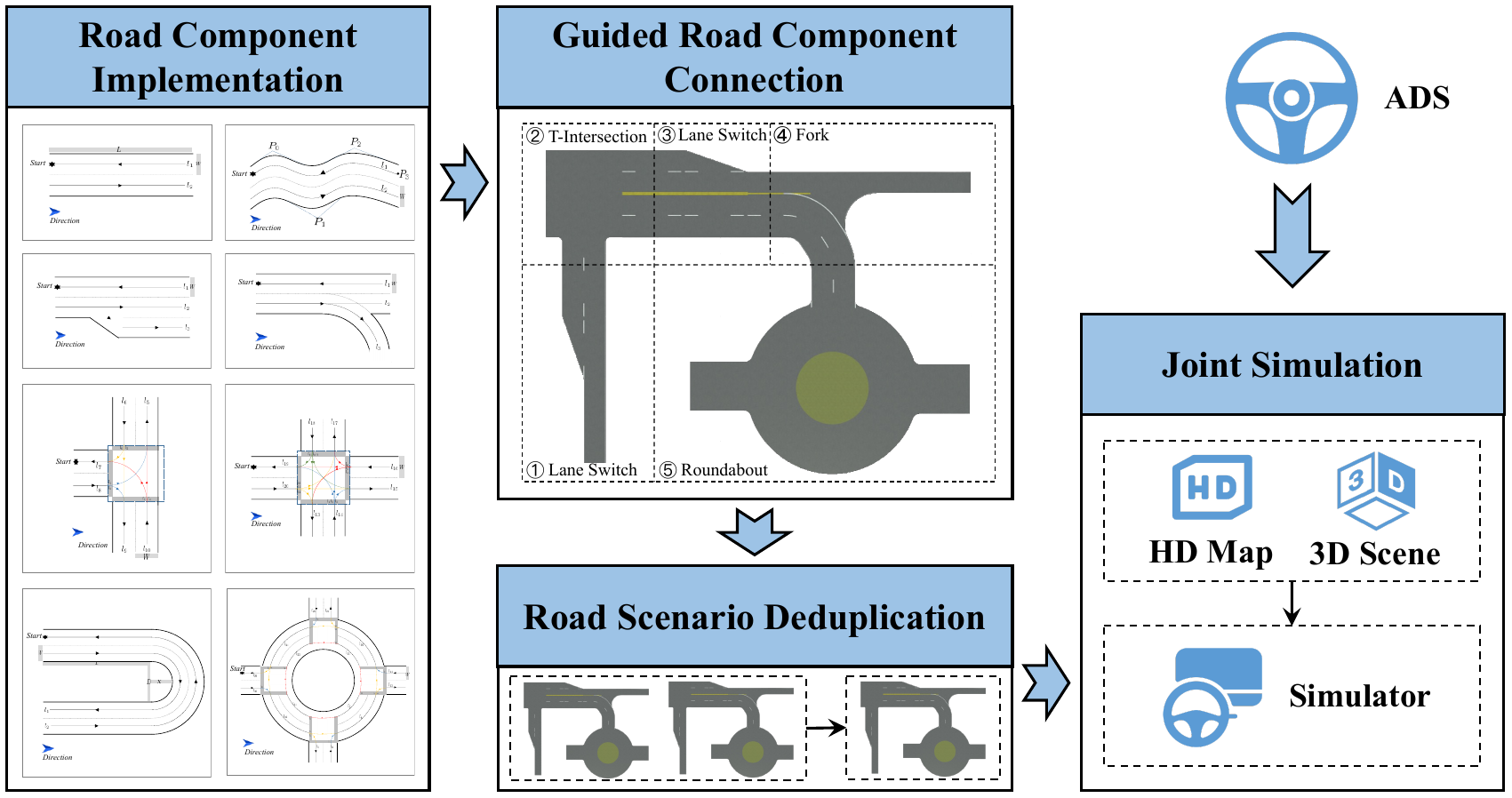}
    \caption{Approach Overview of \tool}
    \label{img:overflow}
\end{figure}

We propose \tool to systematically generate diverse road scenarios. An overview of \tool~is~presented in Fig. \ref{img:overflow}. Each step of \tool is explained below.


\subsection{Road Component Implementation}\label{sec:implement}

\begin{figure}[!t]
\centering  
\subfigure[Straight]{\includegraphics[width=0.42\linewidth]{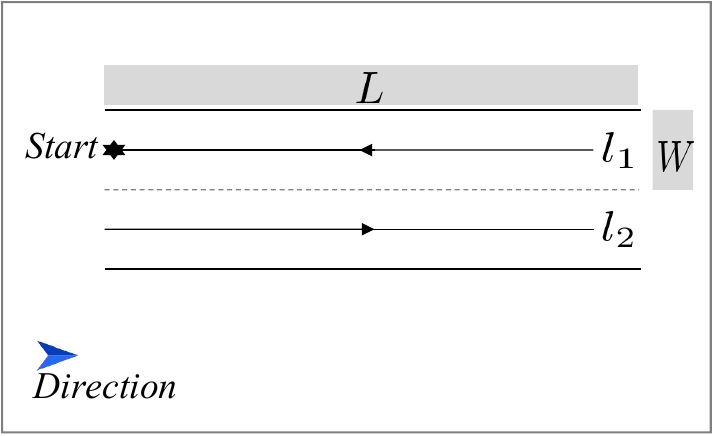}\label{fig:straight}}\hspace{5mm}
\subfigure[Curve]{\includegraphics[width=0.42\linewidth]{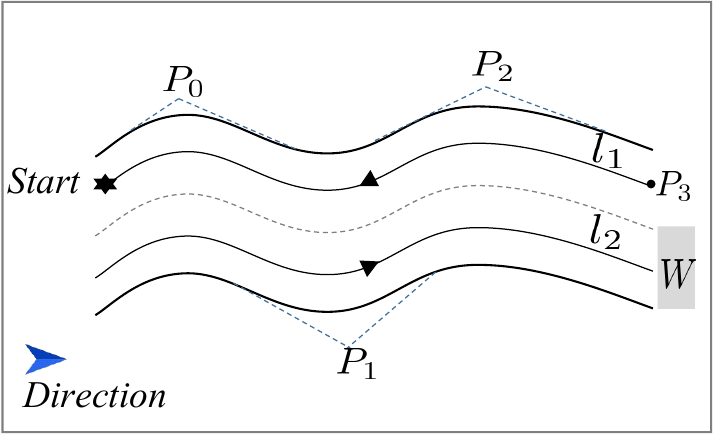}\label{fig:curve}}\hspace{5mm}
\subfigure[Lane Switch]{\includegraphics[width=0.42\linewidth]{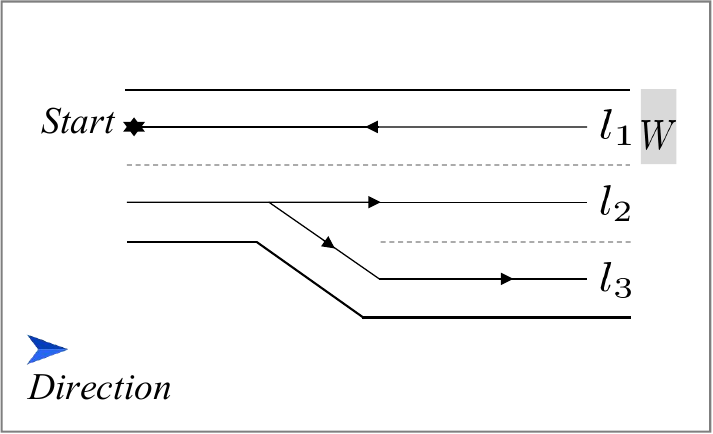}\label{fig:switch}}\hspace{5mm}
\subfigure[Fork]{\includegraphics[width=0.42\linewidth]{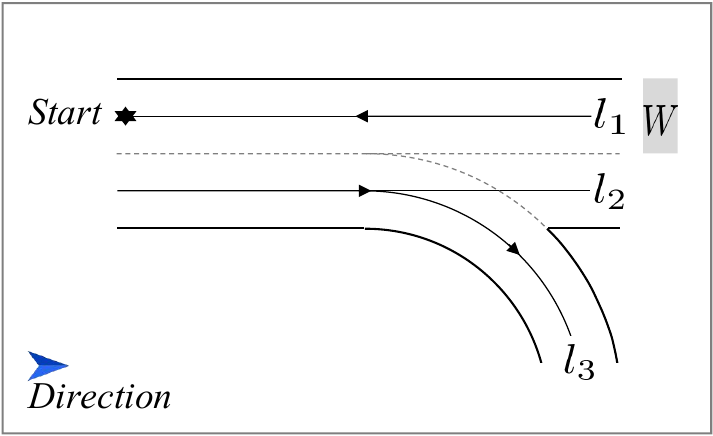}\label{fig:fork}}\hspace{5mm}
\subfigure[T-Intersection]{\includegraphics[width=0.42\linewidth]{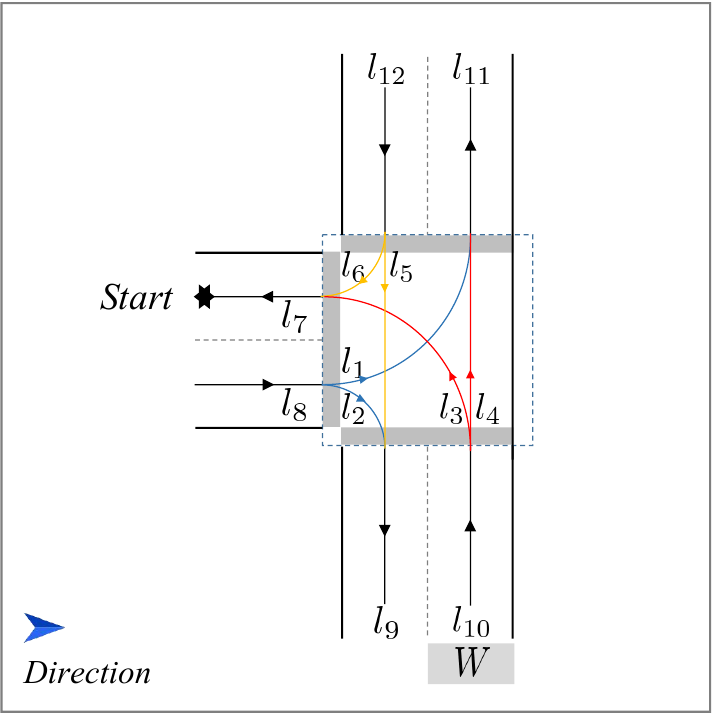}\label{fig:T-intersection}}\hspace{5mm}
\subfigure[Intersection]{\includegraphics[width=0.42\linewidth]{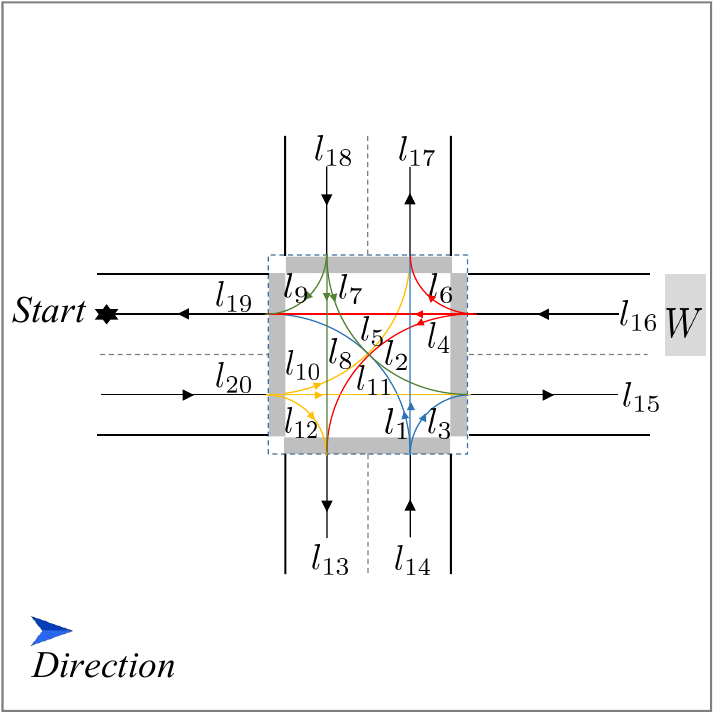}\label{fig:intersection}}\hspace{5mm}
\subfigure[U-Shaped Road]{\includegraphics[width=0.42\linewidth]{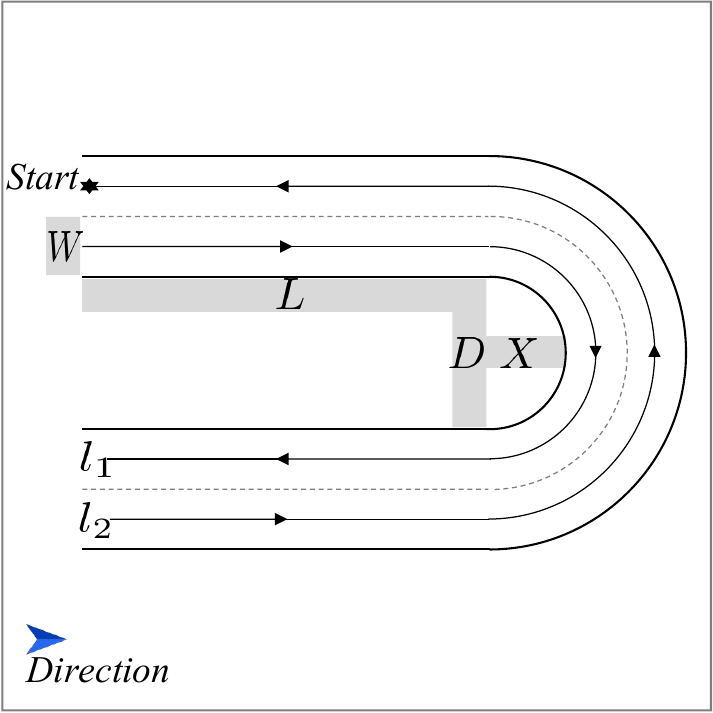}\label{fig:ulane}}\hspace{5mm}
\subfigure[Roundabout]{\includegraphics[width=0.42\linewidth]{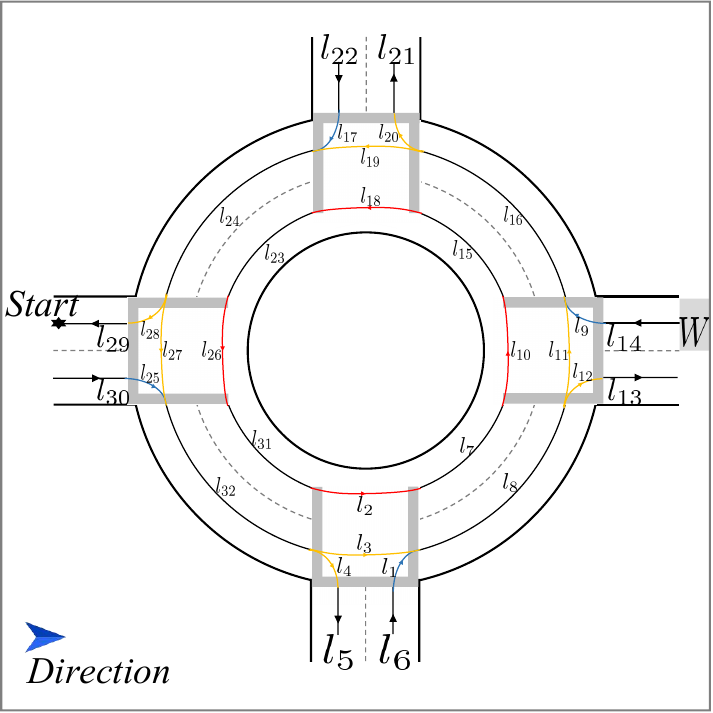}\label{fig:roundabout}}
    \caption{Eight Types of Typical Road Components}
    \label{fig:component}
\end{figure}

Based on our understanding of real-life roads,~we decompose roads into distinct segments, and define eight types of typical road components, as illustrated in Fig.~\ref{fig:component}. To reflect geometry diversity, each road component can be parameterized by road length ($L$), lane width ($W$), the number of lanes ($LaneNum$), the type of lane markings ($LaneMarks$), the coordinate of the starting point ($Start$), and the direction to position the component from the starting point ($Direction$).

 
\begin{itemize}
    \item \textbf{Straight.} As shown in Fig.~\ref{fig:straight}, a straight road follows a linear trajectory without any curves, bends, or turns. 
    
    \item \textbf{Curve.} As shown in Fig.~\ref{fig:curve}, a curve road changes its direction as it progresses. We apply Bézier curves~\cite{mortenson1999mathematics} to construct the lane curve. A Bézier curve $B(t)$ can~be constructed by four control points $P_{0}-P_{3}$, i.e., $B(t)=(1-t)^{3}P_{0}+3(1-t)^{2}tP_{1}+3(1-t)t^{2}P_{2}+t^{3}P_{3},\ t\in [0,1]$. 

    \item \textbf{Lane Switch.} As shown in Fig.~\ref{fig:switch}, a lane switch~road undergoes a transition about the number of lanes (e.g., changing from 2 lanes to 3 lanes). 
    
    \item \textbf{Fork.} As shown in Fig.~\ref{fig:fork}, a fork road is a road~segment where a single road splits into two separate roads, or two separate roads merge into a single road. 
    
    \item \textbf{T-Intersection.} As shown in Fig.~\ref{fig:T-intersection}, a T-intersection is a junction where one road meets another road at a right angle. It creates a three-way intersection where vehicles on the main road can continue straight,~while vehicles on the intersecting road can turn left or right. 
    
    \item \textbf{Intersection.} As shown in Fig.~\ref{fig:intersection}, intersection is a junction where two roads meet or cross each other. 
    
    \item \textbf{U-Shaped Road.} As shown in Fig. \ref{fig:ulane}, a U-shaped road is a road segment where a 180-degree turn~is~required to go in the opposite direction. It can be considered as being composed of two straight road segments and a circular arc segment. 
    
    \item \textbf{Roundabout.} As shown in Fig. \ref{fig:roundabout}, a roundabout~is~a junction where traffic from four roads flows around a central island in a counterclockwise direction~(in~countries with right-hand traffic). 
\end{itemize}

After defining the eight types of road components, we~use Python to implement programming templates for  these road components. These templates can be instantiated to generate MATLAB scripts which describe the road components. As illustrated in Listing~\ref{listing}, all templates are primarily composed of the following steps, i.e., initializing the lanes contained~in a component according to $LaneNum$, drawing coordinates of lanes and boundaries based on $L$, $W$, $Start$ and $Direction$, setting lane markings according to $LaneMarks$, and combining lanes to form the component by assigning references to lane predecessor and successor.

We provide different templates for each road component based on the number of lanes ($LaneNum$) and the type of lane markings ($LaneMarks$). These templates are distinct from each other, encompassing 1 to 6 lanes and any of the seven types of lane markings based on real-world road rules (i.e., white dashed lane lines, white solid lane lines, white double solid lane lines, yellow dashed lane lines, yellow solid lane lines, yellow double solid lane lines, and yellow dashed-solid lane lines). Specifically, we provide a total of 242 unique templates for these eight types of road components. 

When instantiating these templates, we only need~to~specify the road length ($L$), lane width ($W$), coordinate~of~starting point ($Start$), and direction to position the road component ($Direction$). For curve road, we also need to specify $P_{0}-P_{3}$. For U-shaped road, we also need to specify $D$ to denote the distance between the two straight road segments, and $X$ to control the arc segment. 
We can calculate the coordinates of the road and boundaries to fill in the template based on these input parameters, thereby instantiating the component.

\definecolor{codegreen}{rgb}{0,0.4,0}
\definecolor{codegray}{rgb}{0.5,0.5,0.5}
\definecolor{codepurple}{rgb}{0.58,0,0.82}
\definecolor{backcolour}{rgb}{0.95,0.95,0.95}

\lstdefinestyle{mystyle}{
    float=tp,
    backgroundcolor=\color{backcolour},   
    commentstyle=\color{codegreen},
    keywordstyle=\color{magenta},
    numberstyle=\tiny\color{codegray},
    stringstyle=\color{codepurple},
    basicstyle=\ttfamily\fontsize{6.82}{10.5}\selectfont,
    breakatwhitespace=false,         
    breaklines=true,                 
    captionpos=b,                    
    keepspaces=true,                 
    numbers=left,                    
    numbersep=5pt,                  
    showspaces=false,                
    showstringspaces=false,
    showtabs=false,                  
    tabsize=2
}

\begin{figure}
\lstset{style=mystyle}
\begin{lstlisting}[caption=Part of a Sample Template for Road Components,label=listing,language=matlab]
% Initialize lanes in a component using LaneNum.
Lanes(LaneNums) = roadrunner.hdmap.Lane();
% Draw coordinates of lanes and boundaries.
Lanes.Geometry = deal([coordinates of lanes]);
LaneBoundaries.Geometry = deal([coordinates of boundaries]);
% Set lane markings according to LaneMarks.
LaneBoundaries.ParametricAttributes = deal(LaneMarks)
% Combine lanes with predecessors and successors
% where i,j,k represent the ID of lanes.
Lane(i).Predecessors = roadrunner.hdmap.Reference(Lane(j));
Lane(i).Successors = roadrunner.hdmap.Reference(Lane(k));
\end{lstlisting}
\end{figure}

\subsection{Guided Road Component Connection}

    \begin{algorithm}[t!] 
        \caption{Guided Road Scenario Generation}
        \footnotesize
        \label{alg:alg1}
        \LinesNumbered 
        \KwIn{$CompList$, $TotalCount$, $Constraints$, $Candidates$}
        \KwOut{$RoadScenSet$}
        $CompCount$~=~$\emptyset$\;         
        \While{$budget~is~not~reached$}
        {
        $Count$~=~$0;~~CoveredArea$~=~$\emptyset$\;
        $EndpointQueue$~=$~\emptyset;~~RoadScen \Longleftarrow \emptyset$\;
        $C$~=~$selectFirstComp( CompList, CompCount)$\;
        $c$~=~$instFirstComp(Constraints, C $)\;
        $RoadScen \Longleftarrow  c.generateRoad()$\;
        $Count$~++\;
        $CoveredArea$~+=~$c.getCoveredArea()$\;
        $CompCount[C]$~+=~$1$\;
        $EndpointQueue.add(c.getEndpoints()$)\;
        \While{$Count < TotalCount ~\&\&~ EndpointQueue \neq \emptyset$}{
            $p$~=~$EndpointQueue.pop()$\;
            \If {$random()~||~isLast(EndpointQueue,p)$}{
                $Cands$~=~$Candidates[p.Type]$\;
                \While{$length(Cands) > 0$}{
                    $D$~=~$selectLeastUse(Cands, CompCount)$\;
                    $d$~=~$inst(p, Constraints, D, CoveredArea)$\;
                    \If{ $d \neq NULL$ }
                    {
                        $RoadScen \Longleftarrow d.generateRoad()$\;
                        $RoadScen \Longleftarrow connect(p, d)$\;
                        $Count$~++\;
                        $CoveredArea$~+=~$d.getCoveredArea()$\;
                        $CompCount[D]$~+=~$1$\;
                        $EndpointQueue.add(d.getEndpoints());$\\
                        \textbf{break}\;
                        }
                    \Else{
                        $Cands.remove(D)$\;
                    }
                }
    
            }
        }
                $RoadScenSet$~+=~$RoadScen$\;
        }
    \end{algorithm}

We design a guided algorithm to connect instantiated road components to generate diverse road scenarios, as presented in  Algorithm~\ref{alg:alg1}. It has four inputs: (1) $CompList$,~the~242 programming templates for the road components as implemented in Sec.~\ref{sec:implement}; (2) $TotolCount$, the number~of~instantiated road components included in one road scenario; (3) $Constraints$, the constraints about parameters of each road component (e.g., the valid range of road length and~lane width), which need~to be satisfied, when road components~are instantiated, to make the generated road scenarios~as~realistic as possible; and (4) $Candidates$, the candidate road components from the 242 programming templates that~can~be~connected to each type of endpoints. Each road component~has one or multiple endpoints that can be connected to the starting point of other road components. For example,~a~roundabout has three endpoints.~Hence,~we summarize the~types of different endpoints, and compute~their~candidate~road~components. For example, a 2-lane bidirectional solid-line endpoint can be connected to a straight road with 2-lane bidirectional solid-line. The output of Algorithm~\ref{alg:alg1} is a set of road~scenarios $RoadScenSet$ in the form of MATLAB scripts.



This algorithm starts by initializing $CompCount$,~which records the number of times each of the road components~in $CompList$ is used (Line 1). $CompCount$ is used~to~guide our algorithm to favor the selection of least used road components during generation. Then, it generates one road~scenario $RoadScen$ in each loop iteration until certain type of budget is reached, e.g., a time budget of 24 hours is reached (Line 2-35). Specifically, in each iteration, it selects a least used road component $C$ from $CompList$ as the first road component to ensure geometry diversity (Line 5),~and~instantiates~it~to get an instance $c$ of $C$ which satisfies $Constraints$ (Line 6). Then, it updates $Road Scen$, updates $Count$ to record the number of used road components, updates $CoveredArea$ to record~the occupied area of the road scenario, updates $CompCount$~to~record the usage of $C$, and adds the endpoints of $c$ to a queue $EndpointQueue$ (Line 7-11).

As long as the current road scenario can be potentially further connected to other road components, i.e., $Count$ is less than $TotalCount$ and $EndpointQueue$ is not empty (Line 12), it pops from $EndpointQueue$ an endpoint $p$ from which $RoadScen$ is further expanded (Line 13).~If~$random()$~(returning either $true$ or $false$) returns $true$ or $p$ is the last one in $EndpointQueue$ (Line 14), it starts to expand $RoadScen$ at $p$ (Line 15-31). Here, the randomness caused by $random()$ is leveraged to ensure topology diversity.

Then, it obtains from $Candidates$ the candidate road~components $Cands$ that can be connected to $p$ according to the type of $p$ (Line 15). Next, it selects the least used~road~component $D$ from $Cands$ to ensure geometry diversity (Line 17), and instantiates~it~to get an instance $d$ of $D$ which~satisfies $Constraints$, matches with $p$, and does not overlap with the current road scenario (Line 18). Specifically, it uses the covered area of $RoadScen$ and the covered area of $d$, i.e., $CoveredArea$ and $d.getCoveredArea()$, to determine whether overlap occurs. If such a $d$ is found (Line 19),~it~updates $RoadScen$ by adding $d$ and connecting to $d$ through~$p$ (Line 20-21), updates $Count$, $CoveredArea$,~$CompCount$, and $EndpointQueue$ (Line 22-25), and breaks to continue expanding $RoadScen$ at other endpoints (Line 26). If such~a $d$ is not found, it removes $D$ from $Cands$ (Line 29),~and tries to select the next least used road component from $Cands$.


\subsection{Road Scenario Deduplication}

The generated road scenarios can be similar in their topology, i.e., the types of road components and the connections between road components in two road scenarios~can~be~similar, thus hurting the topology diversity. Hence, we propose a similarity metric to measure the topology similarity, and use it to remove duplicated road scenarios to ensure diversity.


We first define a road scenario as an undirected graph in Definition 1 to model the topology of a road scenario.

\begin{theorem}
A generated road scenario can be modeled as an undirected graph $\mathbb{G} = \langle V, E\rangle$, where $V$ is a set~of~vertices denoting the road components within the road scenario, and $E \subseteq V \times V$ is a set of undirected~edges~denoting~the~connections between road components.
\end{theorem}

\begin{theorem}
Given two road scenarios $\mathbb{G}=\langle V,E \rangle$ and $\mathbb{G'}=\langle V', E'\rangle$, and $e = (u, v) \in E$, $u, v \in V$, if $\exists~u', v' \in V'$, $(u', v')\in E'$ such that $Type(u,v)=Type(u',v')$, $e$ is regarded as duplicated in $\mathbb{G'}$, denoted as $DE(e,\mathbb{G'}) = 1$; otherwise, $DE(e,\mathbb{G'}) = 0$. Here, $Type(u, v)$ returns the types of road component $u$ and $v$.
\end{theorem}

\begin{theorem}
Given two road scenarios $\mathbb{G}=\langle V,E\rangle$~and $\mathbb{G'}=\langle V',E'\rangle$, and $u \in V$, if $\forall~(u,v_{i})\in E$, $v_{i} \in V$ such that $DE((u,v_{i}),\mathbb{G'}) = 1$, $u$ and its connections are regarded as duplicated in $\mathbb{G'}$, denoted as $DV(u,\mathbb{G'}) = 1$; otherwise, $DV(u,\mathbb{G'}) = 0$.
\end{theorem}

Based on Definition 2 and 3, we define a similarity metric $Sim_{(\mathbb{G}_1, \mathbb{G}_2)}$ to measure the topology similarity between two road scenarios $\mathbb{G}_1$ and $\mathbb{G}_2$, as formulated in Equation~\ref{eq:1},
\begin{equation}\label{eq:1}
\begin{small}
    \begin{aligned}
        Sim_{(\mathbb{G}_1, \mathbb{G}_2)}=\frac{\sum_{i=1} ^{\parallel V_1 \parallel}{DV(u_i, \mathbb{G}_2)} + \sum_{i=1} ^{\parallel V_2 \parallel}{DV(v_i, \mathbb{G}_1)}}{\parallel V_1 \parallel + \parallel V_2 \parallel}
    \end{aligned}
\end{small}
\end{equation}   
where $\mathbb{G}_1=\langle V_1,E_1\rangle$, $\mathbb{G}_2=\langle V_2,E_2\rangle$, $u_i \in V_1$ and $v_i \in V_2$.


\begin{theorem}
Given two road scenarios $\mathbb{G} =\langle V, E\rangle$ and $\mathbb{G'}=\langle V', E'\rangle$, $\mathbb{G}$ and $\mathbb{G'}$ are regarded as duplicated in the topology if $Sim_{(\mathbb{G},\mathbb{G'})} = 1$.
\end{theorem}

Based on Definition 4, we remove duplicated road scenarios to ensure the topology diversity. Notice that we~can~also use this similarity metric to keep the road scenarios whose similarity to existing road scenarios is below a threshold.

\subsection{Joint Simulation}

Given the deduplicated road scenarios in the form~of~MATLAB scripts, we first adopt MATLAB to compile the scripts into RoadRunner HD map files (i.e., rrhd files). Here we use the rrhd format because RoadRunner provides programmatic interfaces to import rrhd files and export~HD~map~file~types needed by various autonomous driving systems (e.g., Apollo and Autoware) as well as 3D scene file types required~by~various simulators (e.g., SORA-SVL and CARLA). Then,~we use RoadRunner to convert rrhd files into target HD map files and  3D scene files for joint simulation on a target simulator with a target autonomous driving system. 





\section{Evaluation}


We have implemented a prototype of \tool~in~\todo{354K} lines of Python and MATLAB code, and released the~source code of our prototype as well as all the experimental data~at our website \url{https://roadgen.github.io/}.~To evaluate the effectiveness and usefulness of \tool,~we~design the following two research questions (RQs).

\begin{itemize}
    \item \textbf{RQ1 Effectiveness Evaluation}: How is the effectiveness of \tool in generating diverse road scenarios?
    \item \textbf{RQ2 Usefulness Evaluation}: Can the road scenarios generated by \tool be used for simulation?
\end{itemize}

\subsection{Evaluation Setup}

\begin{figure*}[!t]
\centering  
\subfigure[5 Road Components]{\includegraphics[width=0.48\linewidth]{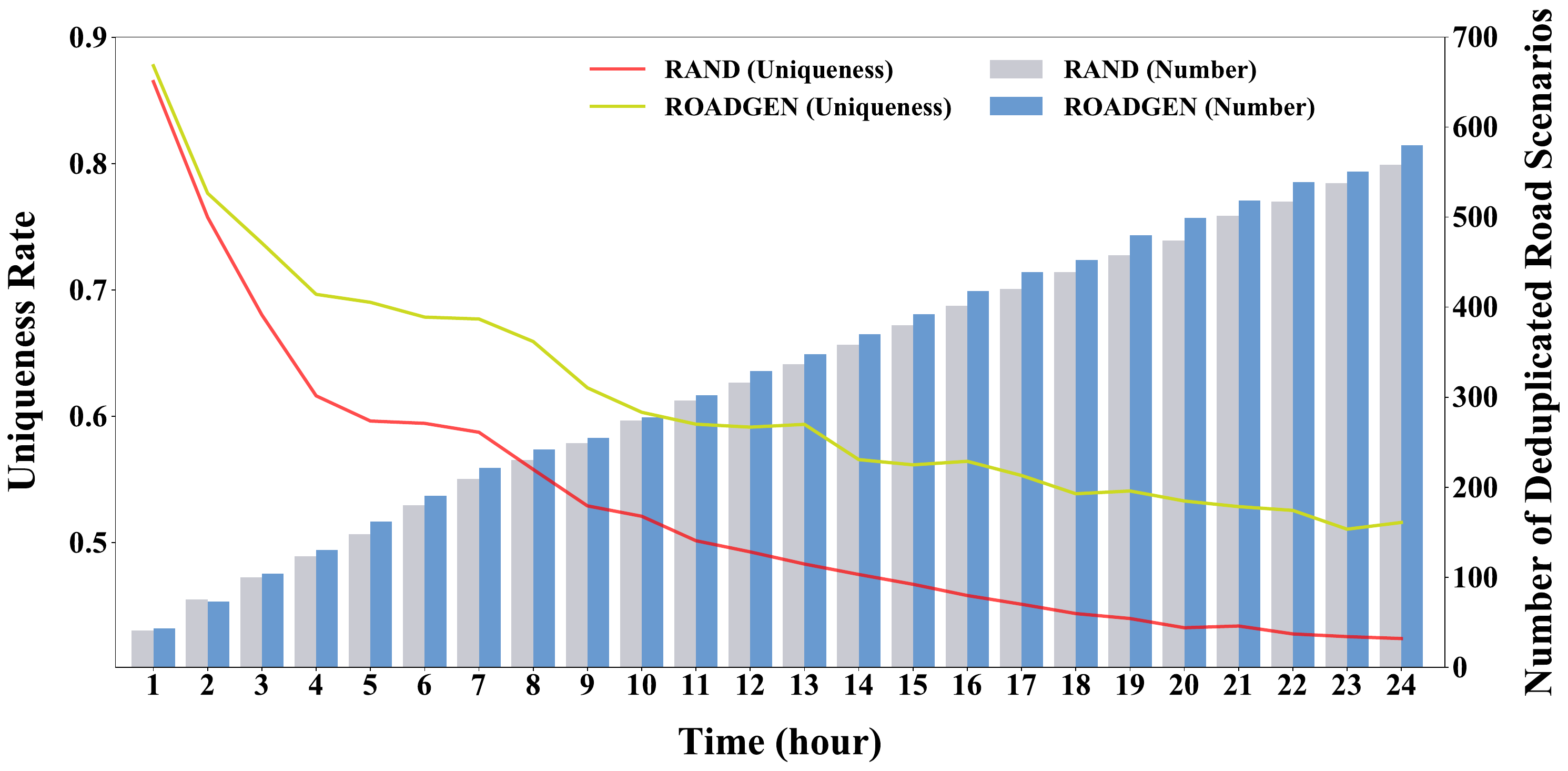}\label{rq2:5w}}
\subfigure[7 Road Components]{\includegraphics[width=0.48\linewidth]{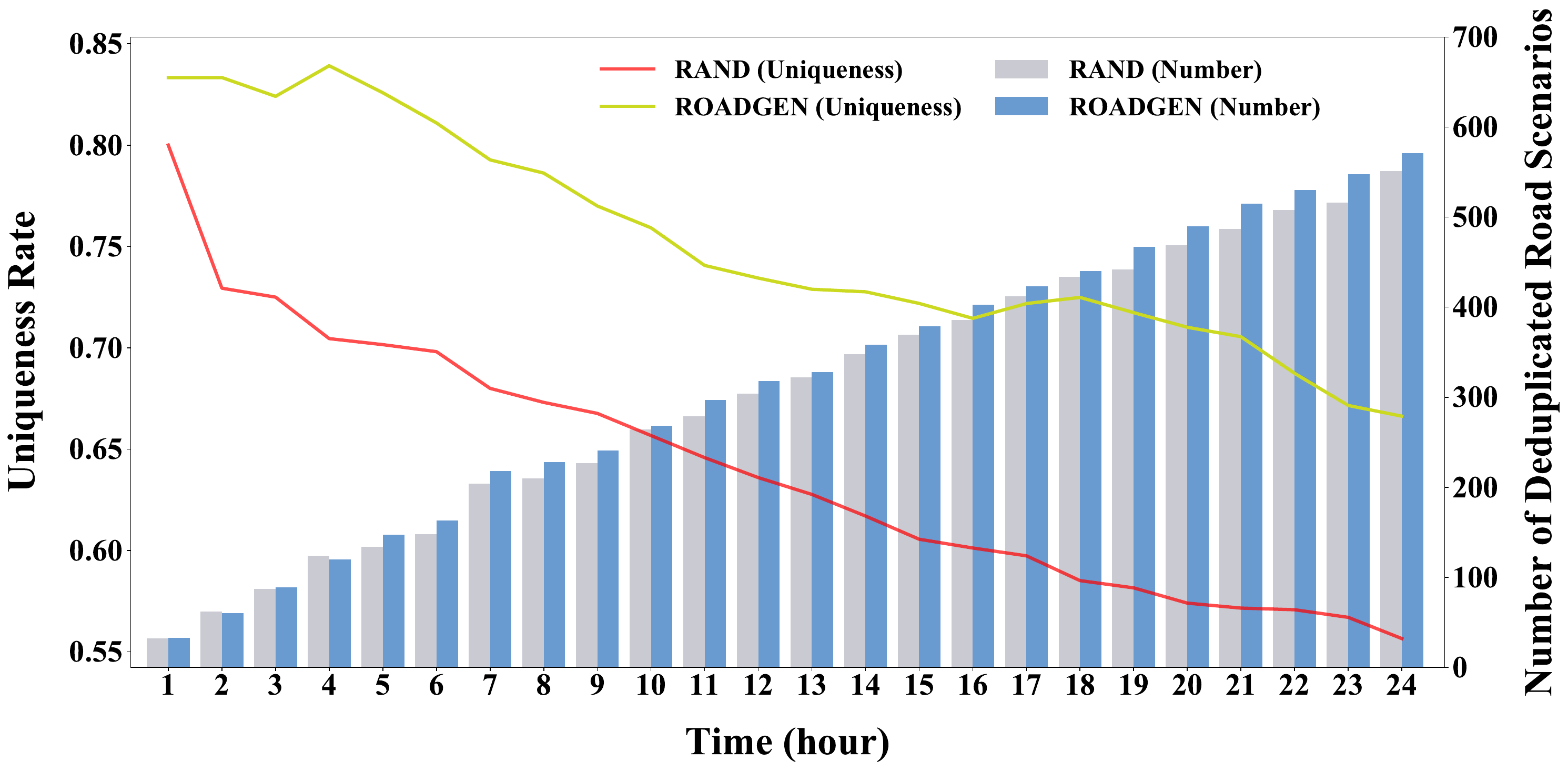}\label{rq2:7w}}
\caption{The Number of Deduplicated Road Scenarios and Uniqueness Rate of \tool and \textsc{Rand} over Time}
\label{fig:sample}
\end{figure*}

\textbf{RQ Setup.} To answer \textbf{RQ1}, we compare the effectiveness of \tool with a baseline approach, referred~to~as~\textsc{Rand}, which randomly selects and connects road components.~We set the number of road components included in each road~scenario to 4, 5, 6, 7 and 8, respectively, and continuously~generate road scenarios for 24 hours using \tool~and~\textsc{Rand}. We record the number of generated road scenarios~after~deduplication, the uniqueness rate (i.e., the rate between~the~number of generated road scenarios after and before deduplication), and the time for covering different road components. We run the experiment 5 times, and report the average results.

To answer \textbf{RQ2}, we leverage Apollo 8.0~\cite{apollo} and SOAR-SVL~\cite{Huai2023} to demonstrate the usefulness of \tool from three perspectives, the usability of 3D scenes, the usability~of HD maps, and the usability for joint simulation.

\textbf{Experimental Environment Setting.} We conduct all the experiments on a Ubuntu 20.04.4 LTS server with 4 NVIDIA GeForce RTX 3090 GPUs, Inter Core i9-10980XE CPU with 3.00GHz processor, and 128GB memory.

\begin{table}[!t]
    \caption{Statistics about Road Scenarios Generated within 24 Hours}
    \label{table1}
    \centering
    \begin{tabular}{ccccc}
        \toprule
        \multirow{2}{*}{Road Scenario Size} & \multicolumn{2}{c}{Number} & \multicolumn{2}{c}{Uniqueness} \\ 
        \cmidrule{2-3}\cmidrule{4-5}
         & \tool & \textsc{Rand} & \tool & \textsc{Rand} \\ 
        \midrule 
        4 Road Components & 400 & 390 & 0.185  & 0.150 \\ 
        5 Road Components & 580 & 558 & 0.516  & 0.424 \\ 
        6 Road Components & 549 & 530 & 0.629  & 0.534 \\ 
        7 Road Components & 571 & 551 & 0.666  & 0.557 \\ 
        8 Road Components & 458 & 432 & 0.602  & 0.527 \\
        \bottomrule 
    \end{tabular}
\end{table}

\subsection{Effectiveness Evaluation (RQ1)}


\textbf{Overall Results.} Table~\ref{table1} presents the statistics about~the road scenarios generated by \tool and \textsc{Rand} within~24 hours. The first column gives the road scenario size in~terms of number of road components, the second and third columns show the number of generated road scenarios~after deduplication, and the fourth and fifth columns list the uniqueness~rate.

Specifically, across the five groups of experiments~with~respect to different road scenario sizes, \tool generate a minimum number of 400 deduplicated road scenarios when road scenario size is set to 4, and a maximum number~of~580 deduplicated road scenarios when road scenario~size~is~set~to 5. In addition, on average, \tool generates \todo{3.9\%} more deduplicated road scenarios than \textsc{Rand} across all the five groups of experiments, which is statistically significant. 


Besides, in terms of uniqueness rate, \tool significantly outperforms \textsc{Rand} by \todo{19.3\%} on average.~In~the four groups of experiments with road scenario size ranging from 5 to 8, the uniqueness rate of \tool fluctuates between 0.5 and 0.7. However, in the group of experiments with road scenario size setting to 4, \tool~exhibits a relatively~low uniqueness rate. This also holds for \textsc{Rand}.~It is because the fewer the number of used road components, the more likely the generated road scenarios will have a higher similarity. 

\textbf{Detailed Results over Time.} Fig. \ref{fig:sample} illustrates a detailed comparison over time between \tool and \textsc{Rand} when road scenario size is set to 5 and 7 in terms of the number of deduplicated road scenarios and the uniqueness rate. Due to space limitation, we provide the results when road scenario size is set to 4, 6 and 8 at our website. 

Specifically, in the early stages of generation, the number of deduplicated road scenarios generated by \tool~is smaller than that of \textsc{Rand}. This is because~\tool~consumes more time than \textsc{Rand} due to our guidance computation, and thus generates a smaller number of road~scenarios before deduplication. However, as time goes by,~the~number of covered road components increases, and \textsc{Rand} gradually generates more duplicated road scenarios, and thus~its~deduplicated road scenarios are gradually less than those generated by \tool. This results in a gradual decline in the uniqueness rate, while the difference between \tool and \textsc{Rand} becomes more significant. 


\begin{figure}[!t]
\includegraphics[width=0.98\linewidth]{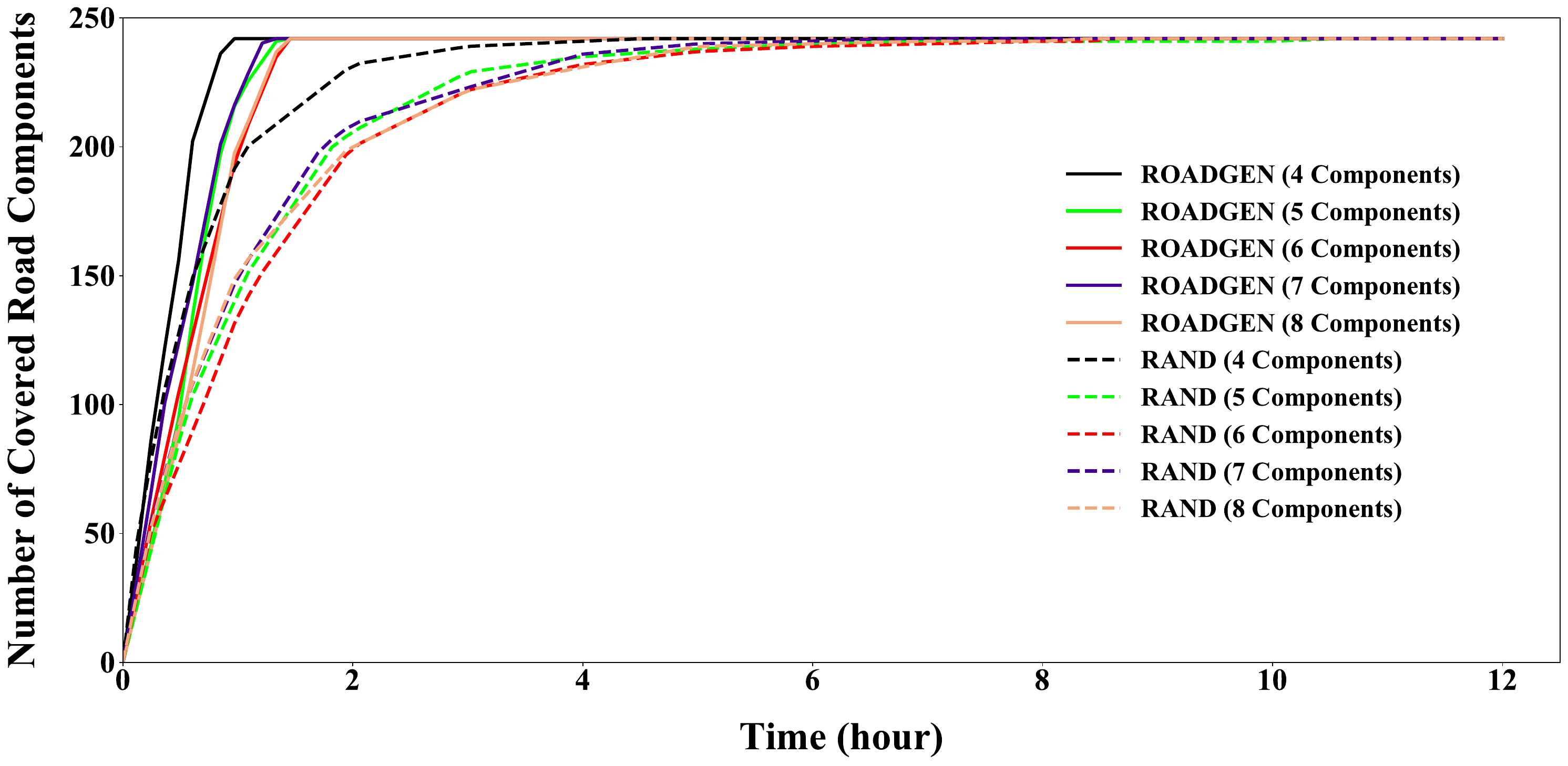}
\caption{Comparison of Time for Covering Different Road Components}
\label{fig:cover}
\end{figure}

\begin{figure*}[!t]
    \centering
    \subfigure[Demonstration of 3D Scenes]{\includegraphics[height=85pt,width=0.32\textwidth]{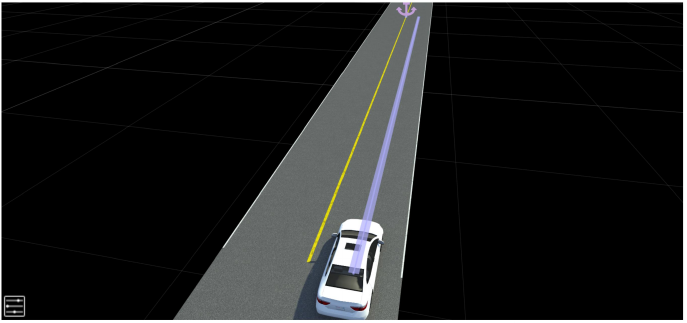}\label{rq3:3Dscenes}}
    \subfigure[Demonstration of HD Maps]{\includegraphics[height=85pt,width=0.32\textwidth]{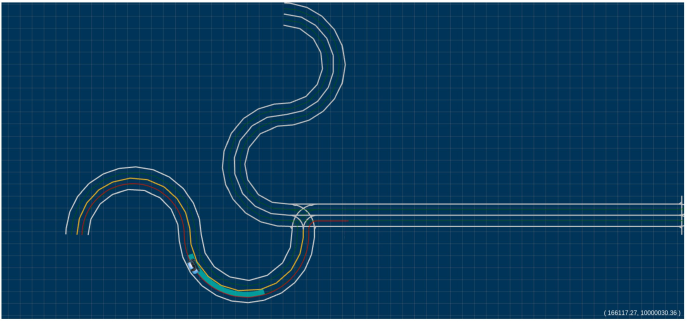}\label{rq3:hdmap}}
    \subfigure[Demonstration of Joint Simulation]{\includegraphics[height=85pt,width=0.32\textwidth]{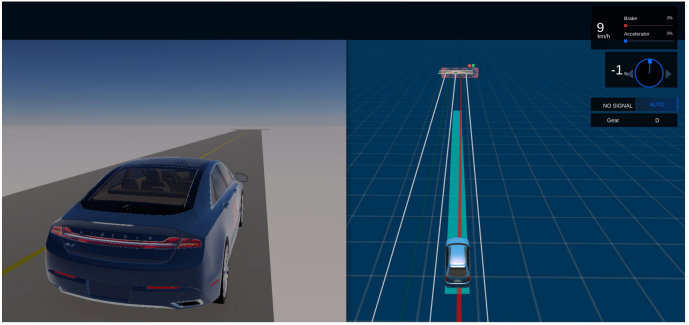}\label{rq3:simulation}}
    \caption{A Case Study that Demonstrates the Usability of 3D Scene Files and HD Maps and the Usability for Joint Simulation}
\end{figure*}

\textbf{Time for Covering Road Components.} Fig. \ref{fig:cover} presents the number of different road components covered by generated road scenarios over time. The solid lines represent the five groups of experiments of \tool, and the dashed lines represent the five groups of experiments of \textsc{Rand}. 

Specifically, \tool respectively spends \todo{0.89}, \todo{1.34}, \todo{1.40}, \todo{1.23} and \todo{1.38} hours to cover all the 242 road components when road scenario size is respectively set to 4, 5, 6, 7 and 8, while \textsc{Rand} respectively costs \todo{4.53}, \todo{10.57}, \todo{8.61}, \todo{6.64} and \todo{8.42} hours. On average, \tool spends \todo{1.24} hours to cover all the 242 road components, while \textsc{Rand} costs \todo{7.75} hours; i.e., \tool is \todo{83.9\%} faster than \textsc{Rand} in covering all the 242 road components.



\textbf{Summary.} These results indicate that our guided~approach \tool is effective in generating more diverse road scenarios in the same time than the baseline random approach, and covers all the different road components more quickly.


\begin{table}[!t]
    \caption{Success Rate of Compiling Scripts into 3D Scene Files}
    \label{table3}
    \centering
    \begin{tabular}{cccc}
        \toprule
       Road Scenario Size &  Total Number & Sample Size & Success Rate\\
        \midrule 
        4 Road Components & 400   & 196 & 96\%\\
        5 Road Components & 580   & 231 & 96\%\\
        6 Road Components & 549    & 226 & 95\%\\
        7 Road Components & 571    & 230 & 93\%\\
        8 Road Components & 458    & 228 & 92\%\\
        \bottomrule
      \end{tabular}
\end{table}

\subsection{Usefulness Evaluation (RQ2)}

Due to the large number of road scenarios in the format of MATLAB scripts generated in \textbf{RQ1}, we evaluate their~usefulness by a sampling approach. Specifically,~we~set~the~confidence level to 95\% and the margin error to 5\% to determine the sample size, and randomly sample road scenarios generated by \tool in \textbf{RQ1}. The third column of Table~\ref{table3} reports the sample size under each road scenario size.

\textbf{Usability of 3D Scenes.} We leverage RoadRunner~to~compile the sampled road scenario scripts into 3D scene files.~The last column of Table~\ref{table3} lists the success rate of compilation. Specifically, more than \todo{92\%} of the road scenarios~scripts~can be successfully compiled into 3D scene files. We investigate the cases where compilation fails, and find that~it~is~caused by the excessive curvature of certain U-shaped road components. We will address this issue~in~future~updates.~Moreover, we validate all 3D scene files using the built-in~scenario~simulation tool in RoadRunner, and the vehicles included therein are able to recognize lane markings, and perform path~planning, lane changes and other behaviors with all 3D scene files. One sample demonstration is illustrated in Fig.~\ref{rq3:3Dscenes}.

\textbf{Usability of HD Maps.} We utilize RoadRunner to convert those sampled road scenario scripts, which are successfully compiled into 3D scene files, into HD map files for the~latest Apollo 8.0. We import these HD map files into Apollo,~and restart Dreamview (i.e., Apollo’s fullstack HMI service) to load the HD maps. We determine the usability of HD maps through manual verification, i.e., running the routing testing in Apollo's built-in sim-control mode, which does not require any third-party simulators. We set points~of~interest~in~sim-control, and let Apollo conduct path planning.~All~HD map files are successfully used in routing testing in sim-control mode. One sample demonstration is shown in Fig.~\ref{rq3:hdmap}.

\textbf{Usability for Joint Simulation.} We first import each 3D scene file into Unity, and manually set the SpawnInfo which contains the starting position and end position~of~a~target~vehicle. Then, we import the resulting scene binary file called AssetBundle into SORA-SVL, and bridge it with Apollo 8.0. We use the SORA-SVL's Python API to start joint simulation in Dreamview and SORA-SVL. All the joint simulations are successful. One sample demonstration is shown in Fig.~\ref{rq3:simulation}. 


\textbf{Summary.} These results indicate that more than \todo{92\%} of the generated road scenarios are useful for joint simulation.


\subsection{Limitations}

While the proposed approach demonstrates promising results, it still suffers several limitations. First, we define eight types of typical road components, which is not meant to be exhaustive but is to illustrate the feasibility of our approach. We plan to further extend the types of road~components~according to regulations in different countries.

Second, we currently only focus on the road-level scenario (i.e., layer 1 of the 5-layer model~\cite{Bagschik2018}) without taking into~account traffic signs, static and dynamic objects, etc.~in~the~upper layers. We are integrating these elements~into~\tool.

Third, we evaluate the usefulness of \tool only on SORA-SVL with Apollo. However, RoadRunner supports~the export of HD map files and 3D scene files in various formats required by various simulators and autonomous driving systems. We believe that \tool is still applicable in other simulators and autonomous driving systems.

\section{Conclusions}

This paper propose \tool to systematically generate diverse road scenarios in both topology and geometry. First, eight types of typical road components are defined~and~implemented. Then, \tool uses a guided~algorithm~to~connect road components to generate road scenarios, and~uses~a~similarity metric to remove duplicated road scenarios.~Our~experimental results have demonstrated the promising effectiveness and usefulness of \tool in generating diverse~road~scenarios for joint simulation. Our dataset of road scenarios is also released for fostering simulation testing. In the future, we plan to extend \tool to support more types of road components and integrate upper-layer scenario elements,~and investigate the applicability of \tool in other simulators and autonomous driving systems.


{\footnotesize
\bibliographystyle{IEEEtranS}
\bibliography{IEEEabrv, src/reference}
}

\end{document}